\documentclass[aps,prx,reprint,superscriptaddress]{revtex4-2}
\usepackage{blindtext}
\usepackage{graphicx}
\usepackage{amsmath}
\usepackage{amssymb}
\usepackage[english]{babel}
\usepackage{xcolor}
\usepackage{siunitx}
\usepackage[colorlinks=true, pdfstartview=FitV, linkcolor=blue,
citecolor=blue, urlcolor=blue]{hyperref}
\usepackage[capitalise]{cleveref}

\usepackage{float}

\makeatletter
\let\saved@includegraphics\includegraphics
\AtBeginDocument{\let\includegraphics\saved@includegraphics}
\renewenvironment*{figure}{\@float{figure}}{\end@float}

\makeatother

\makeatletter
\newcommand\footnoteref[1]{\protected@xdef\@thefnmark{\ref{#1}}\@footnotemark}
\makeatother	

\begin{document}

\title{Dissipation in passive non-reciprocal microwave devices}
\author{Stefano Bosco}
\email{s.bosco@tudelft.nl}
\affiliation{QuTech and Kavli Institute of Nanoscience, Delft University of Technology, Lorentzweg 1, 2628 CJ Delft, Netherlands}

\begin{abstract}
Non-reciprocal devices are key components in both classical and quantum electronics. One approach to realizing passive non-reciprocal microwave devices is through capacitive coupling between external electrodes and materials exhibiting non-reciprocal conductance. In this work, we develop an analytic framework that captures the response of such devices in the presence of dissipation while accounting for the full AC dynamics of the material. Our results yield an effective circuit model that accurately describes the device response in experimentally relevant regimes even at small dissipation levels. Furthermore, our analysis reveals counterpropagating features arising from the intrinsic AC response of the material that could be exploited to dynamically switch the non-reciprocity of the device, opening pathways for tunable non-reciprocal microwave technologies.
\end{abstract}

\maketitle

\section{Introduction}

Non-reciprocal devices are widely used to efficiently route signals and are essential in electronics, microwave photonics, and emerging quantum technologies~\cite{barzanjeh2025nonreciprocity}. Non-reciprocal effects have been proposed and realized in diverse platforms, including magnonics~\cite{PhysRevLett.132.036701,PhysRevApplied.16.064066}, optomechanics~\cite{Fang2017,Barzanjeh2017,PhysRevLett.125.143605}, and heat transport~\cite{PhysRevLett.132.210402}. More recently, superconducting non-reciprocal devices, for example based on the superconducting diode effect~\cite{Ando2020,Nadeem2023,Valentini2024,PhysRevResearch.5.033131,Trahms2023,PhysRevLett.128.037001,PhysRevLett.129.267702,PhysRevLett.128.177001} or multi-terminal devices~\cite{PhysRevLett.130.037001,PhysRevResearch.3.043211,PhysRevResearch.7.013075,Riwar2016,kashuba2024gyrators}, have been suggested as a means to reduce power consumption in classical information processing~\cite{Ingla-Aynes2025,Castellani2025}, as well as to provide new building blocks for quantum information systems~\cite{PhysRevX.11.011032,PhysRevX.13.021017,PhysRevB.99.014514,PhysRevX.15.011072,javed2024utilizing}. Coherent  non-reciprocal devices are being considered as candidates for encoding logical qubits when coupled to superconducting qubits~\cite{PhysRevX.11.011032} and can offer ways to entangle semiconducting qubits over long distances~\cite{PhysRevLett.130.106201,PhysRevB.100.035416,PhysRevLett.133.036301,PhysRevB.96.115407,PhysRevB.93.075301}

In current quantum technologies, low-loss microwave circulators are indispensable for directing the flow of quantum information and suppressing thermal noise~\cite{Arute2019}. State-of-the-art devices are typically passive components that exploit interference effects in combination with magnetic materials~\cite{1125923}. However, achieving microwave operation in the GHz regime restricts these devices to centimeter-scale footprints, which poses challenges for scalability and integration. Active non-reciprocal devices based on reservoir engineering~\cite{PhysRevX.5.021025} and  metamaterials~\cite{PhysRevA.97.043833} have also been explored, but they require continuous external pumping, increasing complexity and heat.
Combining phase delay and switches is also been actively studied as a way to enable circulation~\cite{PhysRevLett.119.147703,PhysRevX.7.041043,PhysRevApplied.11.044048}.

To overcome these limitations, compact passive devices have been proposed that rely on low-loss two-dimensional materials with intrinsic non-reciprocal conductance capacitively coupled to external leads~\cite{PhysRevX.4.021019}. Notable examples include topological materials in the quantum Hall regime~\cite{PhysRevX.7.011007,PhysRevLett.110.016801,Kumada2013,Tarascio}, where non-reciprocal transport emerges under a strong perpendicular magnetic field, as well as anomalous quantum Hall systems that realize similar effects without large external fields~\cite{Mahoney2017,martinez2025circulators,PhysRevB.108.035405,PhysRevB.110.L161403}. These emerging material platforms open pathways to scalable, low-loss, and tunable non-reciprocal microwave technologies.

\begin{figure}
\centering
\includegraphics[width=0.45\textwidth]{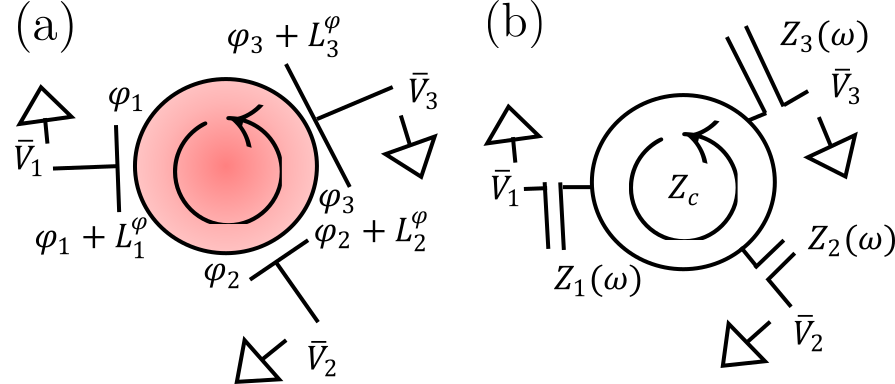}
\caption{\label{fig:fig1} 
\textbf{Device and circuit model.} (a) Sketch of a device comprising a non-reciprocal material capacitively coupled to external electrode of angular length $L_i^\varphi$ with potential $\bar{V}_i$ applied to a common ground. (b) The response of this device can be interpreted via a simple circuit model comprising an ideal circulator with characteristic impedance $Z_c$ coupled to lossy stubs, i.e., transmission lines terminated in open circuits, with frequency dependent impedance $Z_i(\omega)$.  
}
\end{figure}

In this work, we present a general analytical description of this type of passive non-reciprocal devices that, in contrast to existing approaches~\cite{PhysRevX.4.021019,PhysRevApplied.12.014030,PhysRevApplied.7.024030}, fully incorporates the effects of dissipation. We consider the generic architecture sketched in Fig.~\ref{fig:fig1}(a), comprising a material with non-reciprocal conductance capacitively coupled to external driving electrodes. Our model is based on a semiclassical treatment of a Hall material subject to a magnetic field, where both non-reciprocity and losses are captured through the Hall angle $\theta_H$~\cite{Thesis_Bosco}. This parameter interpolates between the reciprocal and dissipative regime ($\theta_H=0$) and the quantum Hall regime ($\theta_H=\pi/2$), where transport is fully non-reciprocal and lossless.

Building on this framework, we derive an effective circuit model, shown in Fig.~\ref{fig:fig1}(b), which accurately captures the device response in the experimentally relevant limits of low losses and low frequencies. By additionally incorporating the intrinsic AC response of the material, our model reveals the presence of counterpropagating resonances. These features provide opportunities to selectively control the propagation direction of signals, offering a new route to dynamically tunable non-reciprocal microwave devices.

\section{Microscopic model}
We consider a semiclassical model of the device in Fig.~\ref{fig:fig1}(a), where the response of a two-dimensional system in the plane $\mathbf{r}=(x,y)$ is governed in the frequency domain by the coupled equations for the excess charge density $\rho$, scalar potential $V$, and current density $\mathbf{j}$~\cite{PhysRevB.32.7676,PhysRevB.95.195317}:
\begin{subequations}
\label{eq:EMP-semiclassical eq}
\begin{align}
\label{eq:Poisson_inv}
V(\textbf{r})&=\bar{V}(\textbf{r})+\int d\textbf{r}' G(\textbf{r},\textbf{r}' )\rho(\textbf{r}') \ , \\
\label{eq:cont}
i\omega\rho(\textbf{r})&=\nabla \cdot \textbf{j}(\textbf{r}) \ , \\
\label{eq:constitutive}
\textbf{j}(\textbf{r})&=-\underline{\sigma}(\textbf{r})\cdot\nabla V(\textbf{r}) \ .
\end{align}
\end{subequations}
Here, Eq.~\eqref{eq:Poisson_inv} represents the inverted three-dimensional Poisson equation with the electrostatic Green’s function $G$ evaluated at the position of the two-dimensional electron gas, including the effect of an external drive $\bar{V}$. Eq.~\eqref{eq:cont} is the continuity equation, while Eq.~\eqref{eq:constitutive} expresses Ohm’s law, relating current density to the electric field through the local conductivity tensor.

When a perpendicular magnetic field $B$ is applied, the Hall effect is captured by an antisymmetric local conductivity tensor. For concreteness, we restrict to a circular device of radius $R$ characterized by
\begin{equation}
\label{eq:conductivity}
\underline{\sigma}(\textbf{r}) =  \sigma_0 n(r) \left(
\begin{array}{cc}
\cos(\theta_H)  & \sin(\theta_H) \\
- \sin(\theta_H)  &  \cos(\theta_H)
\end{array}
\right) \ ,
\end{equation}
and work in cylindrical coordinates $\mathbf{r}=(r,\varphi)$.
To account for spatial inhomogeneity, we introduce a dimensionless function $n(r)$, with support on $r \in [0,R]$, that interpolates from $n(0)=1$ in the bulk of the material to $n(R)=0$ at the edge over a characteristic length scale $l$.

Our results are applicable to the wide class of materials described by Eq.~\eqref{eq:conductivity}, which includes the family of quantum Hall~\cite{girvin2002quantum} and anomalous Hall materials~\cite{RevModPhys.95.011002}. 
To more concretely estimate the AC response and  losses in the material, we consider the AC Drude model. This yields the conductivity scale and Hall angle
\begin{subequations}
\label{eq:drude-cond}
\begin{align}
\sigma_0&=\frac{e^2 n_S}{m^*}\frac{\tau}{\sqrt{(1-i\omega\tau)^2+\omega_c^2\tau^2}} \ , \\
\theta_H&=\text{arctan2}(\omega_c\tau,1-i\omega\tau) \ ,
\end{align}
\end{subequations}
where $n_S$ is the bulk electron density, $\tau$ a characteristic scattering time, $m^*$ the effective mass, and $\omega_c = eB/m^*$ the cyclotron frequency. 

Combining Eqs.~\eqref{eq:cont} and \eqref{eq:constitutive}, we obtain a differential equation relating $\rho$ and $V$:
\begin{equation}
\label{eq:rho-V}
-i\omega \rho = \sigma_0 \Big[ n'\Big(\sin\theta_H \tfrac{\partial_\varphi}{r} + \cos\theta_H \partial_r\Big) + n\cos\theta_H \nabla^2 \Big] V \ .
\end{equation}
This equation reveals that without additional bulk charges, $V$ satisfies the two-dimensional Laplace equation in the bulk of the material and the chiral dynamics of the excess charge along the perimeter is restricted to a small region close to the boundary where $n'=\partial_r n\neq 0$.

Substituting Eq.~\eqref{eq:rho-V} into Eq.~\eqref{eq:Poisson_inv} yields an integro-differential equation for the potential $V$~\cite{Thesis_Bosco}. This equation admits exact solutions in certain limits. In particular, for an infinitely sharp boundary and without external driving, Volkov and Mikhailov showed that a chirally propagating edge charge density (often referred to as edge magneto-plasmons) is redistributed into the bulk by long-range Coulomb interactions, over a length scale set by the kinetic inductance $\text{Im}[\sigma_0 \cos\theta_H]$~\cite{volkov1985theory}. For smoother edges, multiple edge-charge modes can emerge, each propagating with a distinct velocity~\cite{PhysRevLett.72.2935}.

\section{General solution}
\subsection{Driven potential}
The problem simplifies considerably under the local capacitance approximation~\cite{PhysRevB.67.205332}, where the three-dimensional Poisson equation is replaced by the local relation
\begin{equation}
V\approx\bar{V}+\frac{\rho}{c} \ ,
\end{equation}
with capacitance per unit length $c$. 
If we also restrict ourselves to sharp interfaces with $l\to 0$, we find that $\rho\propto n'\approx-\delta(r-R)$ and the bulk potential satisfies the Laplace equation
\begin{equation}
\nabla^2 V=0 \ ,
\end{equation}
complemented by the capacitive boundary condition at $r=R$~\cite{PhysRevX.4.021019}
\begin{equation}
\label{eq:V_LCA}
i\omega (V-\bar{V}(\varphi))= \omega_R \left(\sin\theta_H\partial_\varphi + R\cos\theta_H  \partial_r \right)V \ ,
\end{equation} 
where we introduce the characteristic frequency
\begin{equation}
\label{eq:omegaR}
\omega_R=\frac{\sigma_0}{c R} \ .
\end{equation}
The driving potential $\bar{V}$ is a function of the angle $\varphi$ along the edge and in general it can be decomposed into Fourier harmonics as
\begin{equation}
\bar{V}(\varphi)=\sum_m \bar{v}_m e^{im\varphi}  \ .
\end{equation}

We find an exact analytic solution for this set of equations for arbitrary driving $\bar{V}(\varphi)$ by first considering that the solution of the Laplace equation, non-diverging at $r=0$, is an analytic function of the form
\begin{equation}
\label{eq:Vpot}
V(r,\varphi)=\sum_{m} v_m e^{im\varphi} \left(\frac{r}{R}\right)^{|m|} \ .  
\end{equation}
The coefficients $v_m$ are determined by the boundary condition in Eq.~\eqref{eq:V_LCA} and are given by
\begin{equation}
\label{eq:vm}
v_m=\frac{\omega \bar{v}_m}{\omega-m\omega_R  e^{i (\theta_H - \pi/2) m/|m|}} \ .
\end{equation}

We now turn to a concrete example in which the boundary potential is defined by $N$ electrodes. Each electrode $j$ has angular length $L_j^\varphi$ and extends from $\varphi_j$ to $\varphi_{j+1}=\varphi_j+L_j^\varphi$, with $\varphi_1=0$ and $\varphi_{N+1}=2\pi$. The applied boundary potential can then be written as
\begin{equation}
\bar{V}(\varphi) = \sum_{j=1}^N \big[\Theta(\varphi-\varphi_j)-\Theta(\varphi-\varphi_{j+1})\big]\bar{V}_j \ ,
\end{equation}
where $\bar{V}_j$ is the potential applied to electrode $j$ relative to a common ground. This potential has Fourier coefficients
\begin{equation}
\label{eq:barvm}
\bar{v}_m= -i\sum_{j=1}^N \frac{e^{-i m \varphi_{j}}-e^{-i m \varphi_{j+1}}}{2\pi m} \bar{V}_j\ .
\end{equation}
The potential $V$ is then found by combining Eqs.~\eqref{eq:Vpot}, \eqref{eq:vm}, and~\eqref{eq:barvm}. We note that the summation over Fourier harmonics can be performed exactly.

Here, we assumed the entire perimeter is covered by electrodes and neglect plasmon delays in ungated regions. This approximation is justified in the local capacitance approximation, where interactions in ungated segments are unscreened and plasmon velocities formally diverge~\cite{PhysRevB.95.195317}. More physically, the plasmon delay can be mimicked by smoothing the capacitance profile~\cite{PhysRevX.4.021019} instead of using step functions or by introducing fictitious grounded electrodes in ungated regions~\cite{PhysRevApplied.12.014030}.

\subsection{Microwave response}
To evaluate the device response, we calculate the current $I_i$ collected at electrode $i$ when a potential $\bar{V}_j$ is applied to electrode $j$ while all other electrodes are grounded. The current is obtained by integrating the time-dependent boundary charge density $i\omega \rho$ along the angular extent of electrode $i$. This procedure yields the terminal admittance matrix element
\begin{widetext}
\begin{subequations}
\label{eq:full solution}
\begin{align}
&{Y_{ij}}=\frac{I_{i}}{\bar{V}_j}=i \omega \sigma_0\sum_{m\neq 0} \frac{e^{{i(\theta_H-\frac{\pi}{2})\text{sign}(m)}}\left(e^{i m \varphi_{i}}-e^{i m \varphi_{i+1}}\right)\left(e^{-i m \varphi_{j}}-e^{-i m \varphi_{j+1}}\right)}{2\pi m (\omega-m\omega_R  e^{i (\theta_H - \pi/2)\text{sign}(m)})} \\
&={\sigma_0}\left[
F\!\left(\frac{\omega}{\omega_R},\delta\varphi_{ij},\theta_H\right)+F\!\left(\frac{\omega}{\omega_R},\delta\varphi_{ij}+{L_i^\varphi}-{L_j^\varphi},\theta_H\right)-F\!\left(\frac{\omega}{\omega_R},\delta\varphi_{ij}+{L_i^\varphi},\theta_H\right)-F\!\left(\frac{\omega}{\omega_R},\delta\varphi_{ij}-{L_j^\varphi},\theta_H\right)\right] \ , \\
&F(\omega,\varphi,\theta)=\frac{e^{i \varphi} \Phi\!\left(e^{i \varphi},1,1-\omega e^{-i (\theta-\pi/2)} \right)+\log\!\left(1-e^{i \varphi}\right)}{i2\pi e^{-i (\theta-\pi/2)} } - \frac{e^{-i \varphi} \Phi\!\left(e^{-i \varphi},1,1+\omega e^{i (\theta-\pi/2)} \right)+\log\!\left(1-e^{-i \varphi}\right)}{i2\pi e^{i (\theta-\pi/2)} } \ ,
\end{align}
\end{subequations}
\end{widetext}
where we  introduced the angular distance  $\delta\varphi_{ij}=\varphi_i-\varphi_j$ and $\Phi(a,b,c)$ is the Hurwitz-Lerch trascendent function.

We will now discuss various limits of this general result.

\subsubsection{No Hall effect} 
When $\theta_H=0$, the Hall effect is absent. In this case the material preserves time-reversal symmetry, and the response is fully reciprocal, such that the terminal admittance matrix satisfies $Y_{ij}=Y_{ji}$.

Although Eq.~\eqref{eq:full solution} can be evaluated directly at $\theta_H=0$, further insight can be gained by expanding the response at small frequencies $\omega < |\omega_R|$. In this regime, the function $F$ takes the form
\begin{equation}
F(\omega,\varphi,0)=-\sum_{k=1}^{\infty} \left(i\omega\right)^k \frac{\text{Re}\left[\text{Li}_{1+k}\left(e^{i \varphi}\right)\right]}{\pi} \ ,
\end{equation}
where $\text{Li}_{k}(x)$ is the polylogarithmic function.

This expression shows that $F(0,\varphi,0)=0$, i.e. the system does not transmit DC signals. This behavior is consistent with the capacitive coupling at the boundary, which inherently blocks static responses and effectively decouples all terminals at $\omega=0$. At finite but small $\omega$, the system response acquires both real and imaginary parts, corresponding respectively to dissipative and reactive contributions. Importantly, in this regime the response is non-universal and its precise value depends on the detailed geometry and edge profile of the Hall material~\cite{PhysRevX.4.021019}.

\subsubsection{Quantum Hall limit}
An important regime of the terminal admittance matrix corresponds to the large magnetic field and low dissipation limit, defined by $\omega \ll \omega_c$ and $\omega_c \tau \to \infty$. This regime mimics the quantum Hall effect, where $\theta_H \to \pi/2$, the diagonal elements of the conductivity tensor vanish, and the off-diagonal component reduces to $\sigma_0 = e^2\nu/h$, with filling factor $\nu = h n_S / eB$~\cite{girvin2002quantum}. We restrict ourselves to a positive magnetic field.
In this limit, the plasmonic characteristic frequency becomes purely real-valued, yeilding $\omega_R=e^2\nu/hcR$. 

The exact solution for the terminal admittance matrix takes the form~ \cite{PhysRevApplied.12.014030,Thesis_Bosco}
\begin{subequations}
\begin{align}
\label{eq:YQH}
Y_{ij}=&-\frac{\sigma_0}{2}\!\left[1+i\cot\!\left(\frac{\pi \omega}{\omega_R}\right)\right] \left(1-e^{\frac{i\omega L_i^{\varphi}}{\omega_R}}\right) \\
& \times
\left\{
\begin{array}{lll}
\left(1-e^{\frac{i\omega (2\pi- L_i^{\varphi})}{\omega_R}}\right) & & i=j \\
\left(1-e^{\frac{i\omega L_j^{\varphi}}{\omega_R}}\right) e^{\frac{-i\omega \delta\varphi_{ji}^\circlearrowright}{\omega_R}} & & i\neq j
\end{array} \right. \ ,
\end{align}
\end{subequations}
where $\delta\varphi_{ji}^\circlearrowright$ denotes the angular separation between the right edge of electrode $j$ ($\varphi_{j+1}$) and the left edge of electrode $i$ ($\varphi_{i}$), measured in the clockwise direction.

Since at $\theta_H=\pi/2$ the boundary dynamics are completely decoupled from the bulk potential [see Eq.~\eqref{eq:V_LCA}], this solution is universal and independent of the precise geometry of the quantum Hall material~\cite{PhysRevX.4.021019}.

\subsubsection{Circuit model for small dissipation}
The lossless quantum Hall admittance matrix $Y_{ij}$ in Eq.~\eqref{eq:YQH} admits a simple interpretation in terms of an effective circuit model, see Fig.~\ref{fig:fig1}(b)~\cite{PhysRevApplied.12.014030,Thesis_Bosco}.
The device behaves as an ideal anticlockwise circulator with characteristic impedance
\begin{equation}
Z_c=\frac{1}{2\sigma_0} \ ,
\end{equation}
and scattering matrix 
\begin{equation}
\label{eq:Scircleft}
S_\circlearrowleft =\left(
\begin{array}{ccc}
0 & 1 &0 \\
0 & 0& 1 \\
1 & 0& 0
\end{array}
\right)  \ ,
\end{equation}
which encodes the chirality of plasmon propagation, connected to each external terminal $j$ via a stub impedance
\begin{equation}
\label{eq:stubs}
 Z_{j} (\omega)=\frac{i}{2\sigma_0}\cot\!\left(\frac{\omega L_j^{\varphi}}{2\omega_R}\right) \ ,
\end{equation}
that accounts for the phase delay of plasmons due to their finite propagation velocity.

This circuit model provides an intuitive interpretation of the resonances observed in Hall effect devices. Beyond sharp edges, it also remains accurate for smoother  boundary profiles emerging in electrostatically-defined edges~\cite{PhysRevB.46.4026,PhysRevB.49.8227}. Moreover, it can be systematically generalized to capture slower plasmonic modes~\cite{PhysRevB.95.195317,PhysRevLett.72.2935} by introducing additional equivalent circuits connected in parallel, with appropriately rescaled values of $\sigma_0$ and $\omega_R$ in both $Z_c$ and $Z_j(\omega)$~\cite{PhysRevB.95.195317,Thesis_Bosco}. 

For small deviations of $\theta_H$ from the ideal quantum Hall limit $\pi/2$, the equivalent circuit can be extended to account for dissipation by introducing complex-valued parameters,
\begin{equation}
\label{eq:circuit-model-lossy}
\sigma_0\to \sigma_0 e^{i(\theta_H-\pi/2)\text{sign}(\omega)} \ \ \text{and} \ \ \omega_R\to \omega_R e^{-i(\theta_H-\pi/2)\text{sign}(\omega)} \ .
\end{equation}
This modification incorporates losses both in the plasmon propagation, through the dissipative stubs,
\begin{equation}
 Z_{j} (\omega)=\frac{i e^{-i(\theta_H-\frac{\pi}{2})\text{sign}(\omega)}}{2\sigma_0} \cot\!\left(\frac{\omega L_j^{\varphi}}{2\omega_R}e^{i(\theta_H-\frac{\pi}{2})\text{sign}(\omega)}\right) \ ,
\end{equation} 
and in the circulation itself, via the complex-valued  impedance
\begin{equation}
Z_c=\frac{1}{2\sigma_0} e^{-i(\theta_H-\frac{\pi}{2})\text{sign}(\omega)} \ .
\end{equation}

\begin{figure}
\centering
\includegraphics[width=0.45\textwidth]{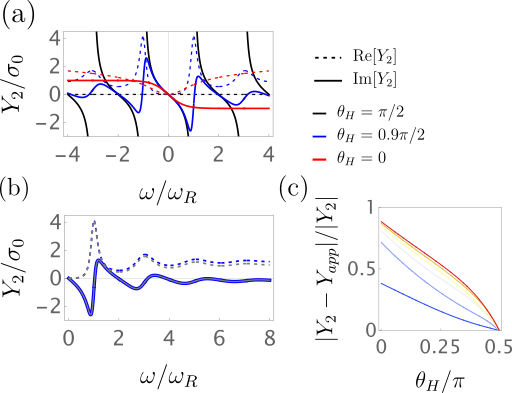}
\caption{\label{fig:Fig2} 
\textbf{Reciprocal dissipative response.} 
(a) Frequency dependence of $Y_2$ for varying Hall angle. We show real and imaginary components with dashed and solid lines. 
In the quantum Hall regime $\theta_H=\pi/2$ (black curves) the system has resonances at $\omega=\omega_R$. These resonances are smoothed by dissipation, when the Hall angle decreases (blue curves), and are completely removed when $\theta_H=0$ (no Hall effect).
(b) For small dissipation ($\theta_H=0.9\pi/2$) and low frequency, the response is accurately captured by our generalized dissipative circuit model (black solid and gray dashed line). (c) The error of this approximation is shown as a function of $\theta_H$ for varying frequency, ranging from $\omega=0.5 \omega_R$ (blue curve) to  $\omega=5.5 \omega_R$ (red curve) and varied with steps of $\omega_R$. 
}
\end{figure}

\section{Microwave devices}

We now turn to concrete devices that serve as illustrative examples of Eq.~\eqref{eq:full solution}. 

\subsection{Reciprocal two-terminal devices}

To interpret the terminal admittance matrix in Eq.~\eqref{eq:full solution}, we begin with the simplest geometry, that is a two-terminal device in which only two electrodes are coupled to the Hall material.
This configuration is fully reciprocal, which can be intuitively understood from the circuit model of Fig.~\ref{fig:fig1}(b): with only two stubs present, no chirality-induced asymmetry arises.

The terminal admittance matrix in this case takes the general form
\begin{equation}
Y=Y_2\left(
\begin{array}{cc}
1 & -1 \\
-1 & 1
\end{array}
\right) \ .
\end{equation}
For symmetric electrodes of equal length $L_1^\varphi=L_2^\varphi=\pi$,  the explicit expression becomes
\begin{widetext}
\begin{equation}
\frac{Y_2}{\sigma_0}=\frac{\Phi \left(-1,1,1+\frac{\omega e^{i (\theta_H-\frac{\pi}{2})}}{\omega_R}  \right)-\log (2)-H_{\frac{\omega e^{i (\theta_H-\frac{\pi}{2})}}{\omega_R}   }}{i\pi  e^{i (\theta_H-\frac{\pi}{2})}}- \frac{\Phi \left(-1,1,1-\frac{\omega e^{-i (\theta_H-\frac{\pi}{2})} }{\omega_R} \right)-\log (2)-H_{-\frac{\omega e^{-i (\theta_H-\frac{\pi}{2})} }{\omega_R}  }}{i\pi  e^{-i (\theta_H-\frac{\pi}{2})}}
 \ ,
\end{equation}
\end{widetext}
where $H_x$ is the $x$th harmonic number.

We now restrict to real-valued $\omega_R$ and $\sigma_0$. 
Figure~\ref{fig:Fig2}(a) shows the real (dashed curves) and imaginary (solid curves) parts of $Y_2$ as functions of frequency, for different values of the Hall angle $\theta_H$.
In the quantum Hall limit ($\theta_H=\pi/2$, black curves), the expression simplifies to~\cite{PhysRevX.4.021019}
\begin{equation}
Y_2(\theta_H=\pi/2)=-i \tan\left(\frac{\pi \omega}{2\omega_R}\right) \ .
\end{equation}
The response is then purely imaginary, with resonances at $\omega=\omega_R(2n+1)$ and zeros at $\omega=\omega_R(2n)$.
This behavior can be directly interpreted from the circuit of Fig.~\ref{fig:fig1}(b): the two-terminal admittance $Y_2=1/Z_1+1/Z_2$ reduces to the series combination of identical stubs $Z_1=Z_2$ [Eq.\eqref{eq:stubs}], which act as short circuits at resonance and open circuits at the zeros.

When small dissipation is included (blue curves), the sharp resonances broaden and their amplitude decreases with increasing frequency. When the Hall angle approaches zero (red curves), these resonant features vanish altogether, reflecting the absence of chiral plasmon propagation along the boundary in this limit.

In Fig.~\ref{fig:Fig2}(b), we compare the exact analytic expression with the approximate dissipative circuit model of Fig.~\ref{fig:fig1}(b), using the substitutions in Eq.~\eqref{eq:circuit-model-lossy}.
For weak dissipation and low frequencies, the lossy stub model reproduces the smoothened resonances with high accuracy, and deviations appear only at higher-order resonances.

Finally, in Fig.~\ref{fig:Fig2}(c) we quantify the error as a function of $\theta_H$ for different frequencies. The results confirm that the lossy stub circuit provides a quantitatively reliable description over a wide range of frequencies, provided dissipation and Hall angle remains small.

\subsection{Non-reciprocal three-port devices}

\begin{figure}
\centering
\includegraphics[width=0.45\textwidth]{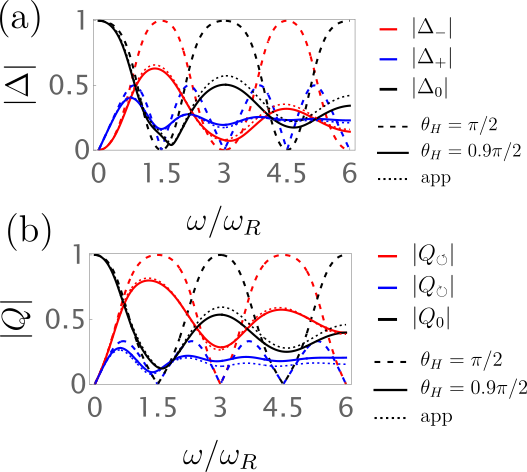}
\caption{\label{fig:fig3} 
\textbf{Non-reciprocal dissipative response.}
We consider a device with three equal terminals impedance-matched with the external electrodes, i.e. $Z_0=1/2\sigma_0$. We operate this device as a two-port gyrator by grounding one electrode in (a) and as a three-port  circulator in (b). We show relevant combinations of the scattering parameters as a function of $\omega$. A finite Hall angle (solid curves) causes dissipation that decreases the amplitude of the transmitted signal as compared to the ideal quantum Hall device (dashed curves). The dampening of resonances is lower at low frequency where is well captured by our approximate dissipative circuit model (dotted curves).
}
\end{figure}

We now turn to the response of non-reciprocal devices, focusing on the three-terminal geometry sketched in Fig.~\ref{fig:fig1}.
For concreteness, we restrict to symmetric electrodes of equal angular length $L_1^\varphi=L_2^\varphi=L_3^\varphi=2\pi/3$, and evaluate Eq.~\eqref{eq:full solution} for this configuration.
Although we emphasize the symmetric three-terminal case, our framework equally applies to multi-terminal devices with $N>3$ as well as asymmetric electrode arrangements, such as the self-matched gyrator proposed in Ref.~\cite{PhysRevApplied.7.024030} and more recently realized in Ref.~\cite{Tarascio}.

\subsubsection{Symmetric gyrator}
We first consider the gyrator~\cite{tellegen1948gyrator}.
An ideal gyrator is a two-port device characterized by the scattering matrix
\begin{equation}
S_G= e^{i\phi}\left(
\begin{array}{cc}
0 & -1 \\
1 & 0
\end{array}
\right) \ .
\end{equation}
where $\phi$ is an arbitrary phase.

Such a device can be implemented from our three-terminal Hall device simply by grounding one of the electrodes, e.g. electrode $3$~\cite{PhysRevApplied.7.024030}.
Starting from the general terminal admittance matrix $Y_{ij}$ in Eq.~\eqref{eq:full solution}, which relates applied potentials to the resulting currents at all electrodes, we obtain a two-port admittance matrix by restricting $Y_{ij}$ to the active electrodes $i,j\in{1,2}$ while taking electrode $3$ as grounded.

From this two-port admittance, the corresponding scattering matrix is obtained through the textbook relation~\cite{pozar2012microwave}
\begin{equation}
\label{eq:Y-to-S}
S = (I+ Z_0 Y)^{-1} (I- Z_0 Y) \ ,
\end{equation}
where $Z_0$ is the characteristic impedance of the external circuit.
In what follows, we assume that the Hall device is impedance-matched, i.e. $Z_0=1/(2\sigma_0)$.
We note that in the quantum Hall regime this condition is difficult to satisfy due to the large value of the resistance quantum ($\nu/\sigma_0\approx 25$~k$\Omega$), but appropriate geometrical modifications can relax this requirement and enable self-matching~\cite{PhysRevApplied.7.024030,Tarascio}.

From the scattering matrix, we define the standard non-reciprocity parameters
\begin{equation}
\Delta_\pm= \frac{S_{12}\pm S_{21}}{2} \ \ \text{and} \ \ \Delta_0= \frac{S_{11}+ S_{22}}{2} \ .
\end{equation}
Here, the gyration parameter $\Delta_-$ quantifies the degree of non-reciprocity, being nonzero only in non-reciprocal devices~\cite{PhysRevApplied.7.024030}.
In contrast, $\Delta_+$ vanishes either for an ideal non-reciprocal device or when the device is fully reflective, i.e. when the relaxation parameter $|\Delta_0|=1$.

Figure~\ref{fig:fig3}(a) shows the frequency dependence of these parameters in the symmetric gyrator.
We compare the dissipationless limit (dashed lines) to the case of finite dissipation (solid lines).
At low frequency, the device is always purely reflective. This behavior also recurs at $\omega = 3n\omega_R$.
In contrast, at $\omega = (3n+1/2) \omega_R$ the device behaves as an ideal gyrator, with perfect non-reciprocity.

This behavior admits a simple circuit interpretation, see Fig.~\ref{fig:fig1}(b).
At $\omega=3n\omega_R$, all three stubs $Z_1=Z_2=Z_3$ behave as open circuits, so the device is reflective.
At $\omega=3(n+1/2)\omega_R$, the stubs instead act as short circuits, yielding an ideal non-reciprocal device.

For finite Hall angle (solid curves), the overall resonance pattern is unchanged, but part of the signal is lost and the amplitude of both transmitted and reflected components is reduced.
The lowest-frequency resonances are less affected by dissipation, while higher-frequency ones are strongly suppressed.
Finally, as shown in Fig.~\ref{fig:fig3}(a), the lossy circuit model obtained using Eq.~\eqref{eq:circuit-model-lossy}, reproduces this behavior accurately (see dotted curves), capturing both the resonance structure and the dissipative smoothing.

\subsubsection{Circulator}
We now consider the three-port circulator~\cite{PhysRevX.4.021019,PhysRevX.7.011007,martinez2025circulators}, which is directly realized in the three-terminal Hall device by measuring the potential at each terminal with respect to a common ground.
From the full three-dimensional terminal admittance matrix, the corresponding scattering matrix can be obtained using Eq.~\eqref{eq:Y-to-S}.
As before, we focus on a matched device with $Z_0=1/2\sigma_0$.

To quantify the circulator’s performance, we define the following parameters:
\begin{subequations}
\label{eq:Qpars}
\begin{align}
Q_\circlearrowleft &= \frac{S_{12}+ S_{23}+S_{31}}{3} \ , \\
Q_\circlearrowright &= \frac{S_{21}+ S_{32}+S_{13}}{3} \ , \\
Q_0 &= \frac{S_{11}+ S_{22}+S_{33}}{3} \ .
\end{align}
\end{subequations}
Here, $Q_\circlearrowleft$ ($Q_\circlearrowright$) equals one for an ideal counterclockwise (clockwise) circulator [see Eq.~\eqref{eq:Scircleft}], while $Q_0$ quantifies reflection, reaching unity only for a fully reflective device.

Figure~\ref{fig:fig3}(b) shows these parameters as functions of frequency.
The behavior mirrors that of the gyrator: the device is fully reflective at $\omega = 3n\omega_R$, while ideal non-reciprocal circulation is achieved at $\omega = 3(n+1/2)\omega_R$.
For finite Hall angles, the amplitude of the circulating signals decreases, reflecting the dissipative losses, but the overall resonance pattern remains unchanged.
As with the gyrator, this dissipative behavior is well captured by our lossy circuit model.

\subsection{AC response of material}

\begin{figure*}
\centering
\includegraphics[width=0.9\textwidth]{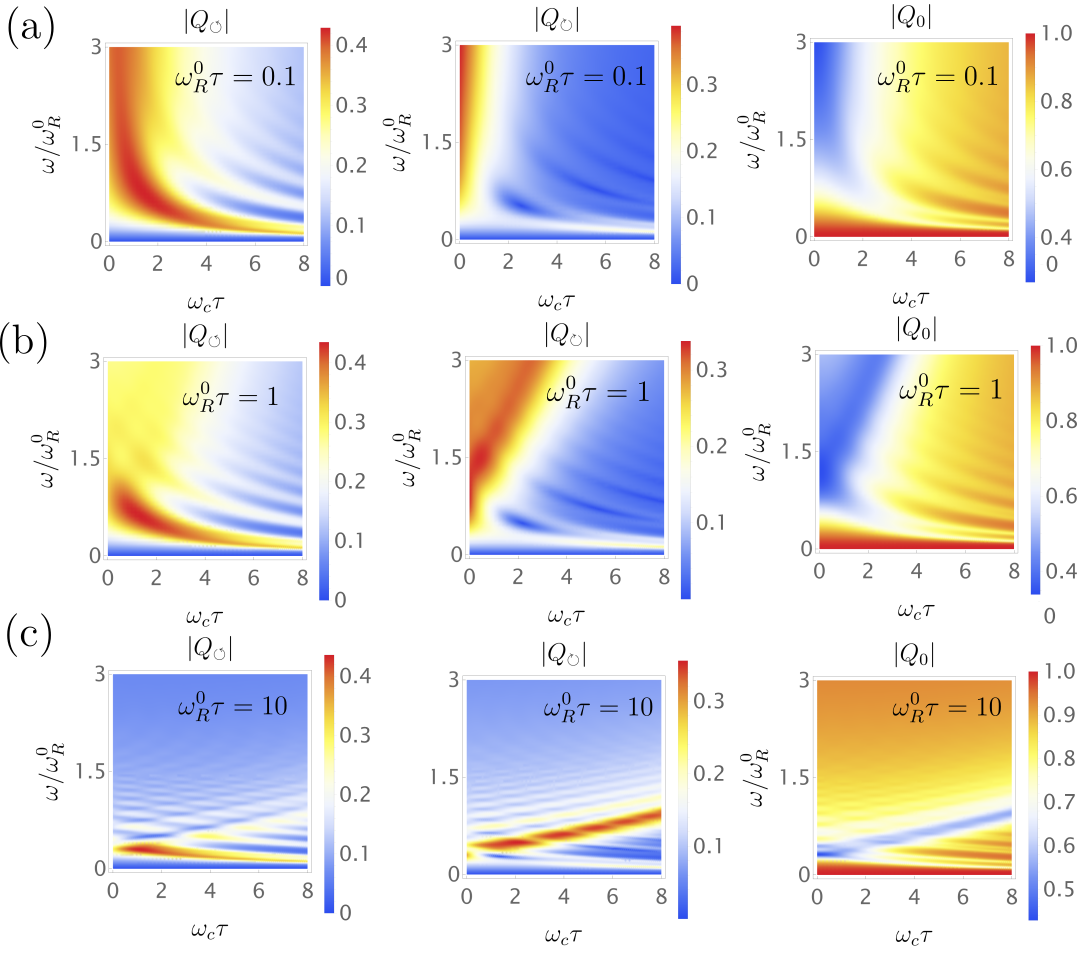}
\caption{\label{fig:fig4} 
\textbf{Intrinsic AC response.}
We consider the response of a three-port dissipative circulator, and show  relevant combinations of scattering parameters against frequency ($\omega/\omega_R^0$) and magnetic field ($\omega_c \tau \propto B$).
To capture the effect of an increasing magnetic field, the external electrodes are impedance-matched with the zero-frequency and zero-field characteristic impedance of the device, i.e. $Z_0=1/2 \sigma_0(\omega=\omega_c=0) $. In analogy, we normalize $\omega$ over $\omega_R^0=\sigma_0(\omega=\omega_c=0)/cR$, and the Hall angle is varied as $\theta_H=\text{arctan2}(\omega_c\tau, 1-\omega\tau) $.
In rows (a), (b), and (c), we show the effect of an increasing value of the kinetic inductance of the device, parametrized by an increasing the product $\omega_R^0\tau$.
Row (a) shows the response for negligible kinetic inductance, where we identify the typical $\propto 1/B$ resonance lines.
As we increase the kinetic inductance, in row (b), additional counter rotating resonances $\propto B$ begin appear, that become sharp resonances at large kinetic inductance in row (c).
}
\end{figure*}

Our general model, and in particular the terminal admittance matrix in Eq.~ \eqref{eq:full solution}, not only captures dissipation but also includes the intrinsic AC response of the material.
This originates from the fact that the conductivity tensor generally has a reactive component, for example due to kinetic inductance, leading to complex-valued $\sigma_0$ and $\theta_H$ [see Eq.~\eqref{eq:drude-cond}].
In our non-reciprocal device this modifies $\omega_R \propto \sigma_0$, which itself becomes complex [see Eq.~\eqref{eq:omegaR}].

In Fig.~\ref{fig:fig4}, we analyze how the intrinsic AC response modifies the operation of the three-port symmetric circulator.
Specifically, we plot the $Q$ parameters defined in Eq.~\eqref{eq:Qpars} as functions of both the frequency $\omega$ and the cyclotron frequency $\omega_c \propto B$, normalized by the scattering time $\tau$.
To capture the full dependence on the magnetic field $B$, we normalize frequencies by $\omega_R^0$, defined as $\omega_R$ evaluated at $\sigma_0(\omega=\omega_c=0)$.
Similarly, we assume the external circuit to be matched to $Z_0=1/2\sigma_0(\omega=\omega_c=0)$.

The intrinsic AC response of the material is encoded in the frequency dependence of $\sigma_0(\omega)$ and $\theta_H(\omega)$, both of which depend on the scattering time $\tau$.
At low frequencies, this dependence results in a kinetic inductance that renders the diagonal components of the conductivity tensor complex, i.e. $\sigma_{xx}\approx\sigma_{xx}(1-i\omega\tau)$~\cite{volkov1985theory}.
To explore this effect, we vary $\tau$ normalized by $\omega_R^0$.

In Fig.~\ref{fig:fig4}(a), we focus on the small kinetic inductance case. Here, we recover the expected $\omega_R\propto \sigma_0\propto 1/B$ scaling of the resonance lines~\cite{PhysRevX.7.011007}. Here, we neglect the flattening of the resonances caused by quantum Hall plateaus.
Dissipation suppresses higher-order resonances, leaving only the lowest ones visible. These results are consistent with previous numerical results for static dissipation in Hall devices~\cite{Placke2017}.

At intermediate kinetic inductance, $\tau \sim \omega_R^0$ in Fig.~\ref{fig:fig4}(b), we observe a smoothening of the low-frequency, low-field counterclockwise resonance. At the same time, more resonant features begin to emerge in the clockwise circulation and in the reflection response.

At large kinetic inductance, in Fig.~\ref{fig:fig4}(c), these features evolve into sharp resonances with an opposite (clockwise) circulation compared to the expected chirality of the plasmons.
Remarkably, these counter-circulating resonances scale $\propto B$, in contrast to the $1/B$ scaling of the direct plasmonic resonances.

We emphasize that while extrinsic circuit elements (such as parasitic capacitances) can also generate counter-circulating signals~\cite{PhysRevX.7.011007, Thesis_Bosco}, they typically retain the same $1/B$ scaling as the primary plasmonic resonances.
The distinct $B$-scaling observed here clearly distinguishes these additional resonances arising from the intrinsic reactive response of the material.

Importantly, even in the lossy regime these counter-resonances remain as strong as the direct circulating features.
This suggests a potentially practical route to selectively control the direction of signal propagation in Hall-effect based circulators.

\section{Conclusion}
In summary, we have developed a general analytic framework describing the microwave response of passive non-reciprocal devices, which naturally incorporates dissipation, geometric effects, and the intrinsic AC response of the underlying material.
Our analytic solution provides clear physical interpretation of the system response in terms of a lossy stub circuit model, which we show accurately captures the device response in the experimentally relevant regime.
Furthermore, by extending the model to include frequency-dependent conductivities, we revealed features, such as counter-circulating resonances with distinct magnetic field scaling, that are not captured by extrinsic circuit elements.
Altogether, our results establish a unified description of microwave devices based on non-reciprocal materials, bridging microscopic material response and macroscopic circuit performance, and providing practical guidelines for designing next-generation compact non-reciprocal components.

\section{Acknowledgments}
We thank David DiVincenzo for stimulating discussions. 
We are also grateful to Aldo Tarascio, Rafael Eggli, Miguel Carballido, Yiqi Zhao, and Dominik Zumb{\"u}hl for constant insights and for sharing experimental data prior publication. This work was supported by NCCR Spin (grant number 225153).

\bibliography{references}
\end{document}